\documentclass[12pt]{article}

\usepackage{amssymb}
\usepackage{amsmath}
\usepackage{graphicx}
\usepackage{subfigure}
\usepackage{textcomp}
\usepackage{color}
\usepackage{amsfonts}
\usepackage{bbold}
\usepackage{dsfont}
\usepackage{hyperref}
\usepackage{setspace}

\begin{document}

\centerline{\large{\bf{Electron doping and magnetic moment formation in N- and C-doped MgO}}}
\vspace{0.5cm}
\centerline{A. Droghetti and S. Sanvito}
\centerline{\it{School of Physics and CRANN, Trinity College, Dublin 2, Ireland}}

\begin{abstract}

The formation of the magnetic moment in C- and N-doped MgO is the result of a delicate interplay between Hund's 
coupling, hybridization and Jahn-Teller distortion. The balance depends on a number of environmental variables 
including electron doping. We investigate such a dependence by self-interaction corrected density functional theory 
and we find that the moment formation is robust with respect to electron doping. In contrast, the local 
symmetry around the dopant is more fragile and different geometries can be stabilized. Crucially the magnetic 
moment is always extremely localized, making any carrier mediated picture of magnetism in $d^0$ magnets unlikely.
\end{abstract}

In the last few years an intensive research effort has been devoted to synthesize diluted magnetic semiconductors (DMS)
in the hope of finding a material with semiconducting properties and ferromagnetism at room temperature (RT). Standard DMS 
are produced by doping conventional semiconductors with transition metals \cite{1}, and the magnetic coupling 
usually originates from some carrier-induced mechanism. More recently two main experimental facts have challenged
this ``traditional'' picture. Firstly magnetism was claimed in oxides DMS on both sides of the metal insulator
transition \cite{Behan}, indicating that carriers alone are not sufficient to explain the magnetism and that intrinsic defects
play a crucial r\^ole \cite{DasZnO}. Secondly, defect-rich or intentionally $p$-doped oxides revealed the possible evidence 
for RT magnetism. This second class of phenomena has been named $d^0$ magnetism, indicating that no ions with partially filled 
$d$-shells are at the origin of the magnetic moment. The examples of $d^0$ magnets are many and include thin films of HfO$_2$\cite{4}, 
TiO$_2$\cite{5,6,7}, In$_2$O$_3$\cite{8}, C-doped ZnO\cite{2,3}, and nanoparticles of different materials\cite{16}. Furthermore 
studies over chemically synthesized powders suggest that $d^0$ magnetism can be found in bulk materials and not 
only in surfaces\cite{9,10}.

Usually the formation of the magnetic moment is explained in terms of spin-polarized holes localized either at cation vacancies
molecular orbitals \cite{11,13} or at the $p$-orbitals of the doping impurity\cite{13,12,14,15}. However,
a clear understanding of the driving mechanism behind $d^0$ ferromagnetism is still unavailable. 
In order to understand its uniqueness we recall that the record Curie temperature ($T_\mathrm{C}$) 
for GaAs:Mn (the DMS prototype) is 200~K\cite{17}, obtained for $8~\%$ Mn doping and the largest hole concentration 
achievable. Since cation vacancies in oxides hardly reach concentrations exceeding $1~\%$ and both C and N cannot be
doped abundantly \cite{3}, one can conclude that a similar $T_\mathrm{C}$ in $d^0$ magnets would require a magnetic 
interaction around 10 times stronger than that between Mn in GaAs:Mn. Considering that our argument 
neglects any considerations about percolation and the fact that even the origin of the magnetic moment is not established 
with certainty, it is fair to say that any claim of $d^0$ ferromagnetism should be considered exceptional. 

In this arena almost all the theoretical predictions are based on density functional theory (DFT) using either the local spin 
density (LSDA) or the generalized gradient approximation (GGA). These describe the ground state of $d^0$ magnets 
as ferromagnetic and metallic, and in fact most of the time as half metallic. However strong electron correlations may play a 
fundamental r\^ole in the magnetic moment formation, which subtly depends on the interplay between covalency, Hund's coupling 
and polaronic distortion around the impurity \cite{13}. Corrections to the LSDA/GGA such as the LDA+$U$ or the self-interaction 
correction (SIC) schemes often return an insulating ground state and no long range magnetism. For instance in rock-salt oxides 
(MgO, CaO and SrO) doped with N substituting for O \cite{13,14,15}, one finds that the extra hole entirely localizes around one 
of the $2p$ orbitals as a consequence of the coupling with phononic modes. Thus the physics of these materials reminds that of 
the manganites \cite{18}, with the difference that in $d^0$ magnets the moment is associated to the 2$p$ atomic shell. 

In analogy with the manganites, one expects that the interplay between charge, spin and orbital degrees of freedom 
may lead to a number of cooperative physical phenomena\cite{18}. In particular the long range magnetic order in $d^0$ 
magnets is intimately related to the formation of the moment itself \cite{12}, with Stoner and spin-wave excitations 
probably competing. Thus both the moment and the magnetic coupling becomes sensitive to environmental 
variables such as doping, charge fluctuations and temperature. 
In this work we investigate the effects of one of these variables, namely electron doping, over the magnetic moment formation
in C- and N-doped MgO. In particular we will answer two fundamental questions: 1) does the magnetic moment survive to 
electron doping? and 2) what is the response of the lattice to such an electron doping?

Our calculations are performed by using a development version of the DFT code {\sc siesta}\cite{19}, implementing the atomic SIC 
scheme (ASIC)\cite{20}. The core electrons are treated with norm-conserving Troullier-Martin pseudopotentials and the valence charge 
density is expanded over a numerical orbital basis set, including double-$\zeta$ and polarized functions \cite{19}. 
The real space grid has an equivalent cutoff larger than 800~Ry. Calculations are performed with supercells of 96 atoms including $k$-point 
sampling over at least 25 points in the Brillouin zone. Atomic coordinates are relaxed by conjugate gradient until the forces 
are smaller than 0.01~eV/\AA.

As already mentioned when N or C replace O (N$_\mathrm{O}$ and C$_\mathrm{O}$) the MgO local cubic symmetry 
is reduced\cite{13}. In both cases the bonds with Mg are longer than the Mg-O one as a consequence of the different ionic radii. 
However, for N$_\mathrm{O}$ two of the three $p$-bonds contract and the hole localizes around the remaining long one (N$_\mathrm{O}$
configuration), while for C$_\mathrm{O}$ only one bond contracts and the two holes localize around the remaining two long 
bonds (C$_\mathrm{O}$ configuration). In both cases the system is insulating with either one (C$_\mathrm{O}$) or two 
(N$_\mathrm{O}$) fully filled $p$ orbitals. We now consider C$_\mathrm{O}$ and add a fractional charge $\Delta n$ to the 
supercell, with charge neutrality ensured by a compensating positive background. One then expects that the electronic structure 
and the relaxation becomes progressively similar to that of N$_\mathrm{O}$. 

Our results are presented in Tab.~\ref{Tab1}, where we list the Mg-C bond lengths, $d_\mathrm{Mg-C}$, and the cell 
magnetic moment, $\mu$, as a function of $\Delta n$.
\begin{table}[h]
\centerline{\begin{tabular}{cccccccc} \hline\hline
$\Delta n$ ($-e$)  & $d_\mathrm{Mg-C}$ (\AA) [C$_\mathrm{O}$-Rel] & $d_\mathrm{Mg-C}$ (\AA) [N$_\mathrm{O}$-Rel] & $\mu$ ($\mu_\mathrm{B}$)\\ \hline\hline
& & &\\0.0 &2.176 (4), 2.152 (2) &2.176 (4), 2.152 (2) & 2.0 & \\ & & &\\
0.2        &2.163 (4), 2.146 (2) &2.170 (2), 2.161 (2), 2.143 (2) &1.8 & \\& & & \\
0.4        &2.149 (4), 2.136 (2) &2.157 (2), 2.143 (2), 2.131 (2) &1.6 & \\& & & \\
0.6        &2.135 (4), 2.120 (2) &2.141 (2), 2.126 (2), 2.120 (2) &1.4 & \\& & & \\
0.8        &2.120 (4), 2.110 (2) &2.122 (2), 2.116 (2), 2.112 (2) &1.2 & \\& & & \\
1.0        &2.105 (4), 2.099 (2) &2.105 (2), 2.099 (4) &1.0 & \\& & & \\
\hline\hline
\end{tabular}}
\caption{\label{Tab1} Mg-C bond length, $d_\mathrm{Mg-C}$, and supercell magnetic moment $\mu$, of C$_\mathrm{O}$
as a function of the electron doping $\Delta n$. We indicate as [C$_\mathrm{O}$-Rel] the relaxed structure obtained from 
an atomic relaxation initiated at the $\Delta n$=0 C$_\mathrm{O}$ geometry, and as [N$_\mathrm{O}$-Rel] that initiated at the
$\Delta n$=0 N$_\mathrm{O}$ geometry. In the bracket beside the bond length we indicate the degeneracy of the particular 
bond.}
\end{table}
Importantly all the structural calculations turned out to be sensibly dependent on the initial conditions for the relaxation. In particular we find
that all the relaxations initialized at the $\Delta n$=0 C$_\mathrm{O}$ configuration converged to a local geometry presenting 4 long
bonds and 2 short ones, i.e. still presenting the symmetry of C$_\mathrm{O}$. In contrast those initialized at the $\Delta n$=0 
N$_\mathrm{O}$ configuration converged to a local geometry with two long, two short and two intermediate bonds. 

\begin{figure}[ht]
\centerline{\includegraphics[scale=0.26,clip=true]{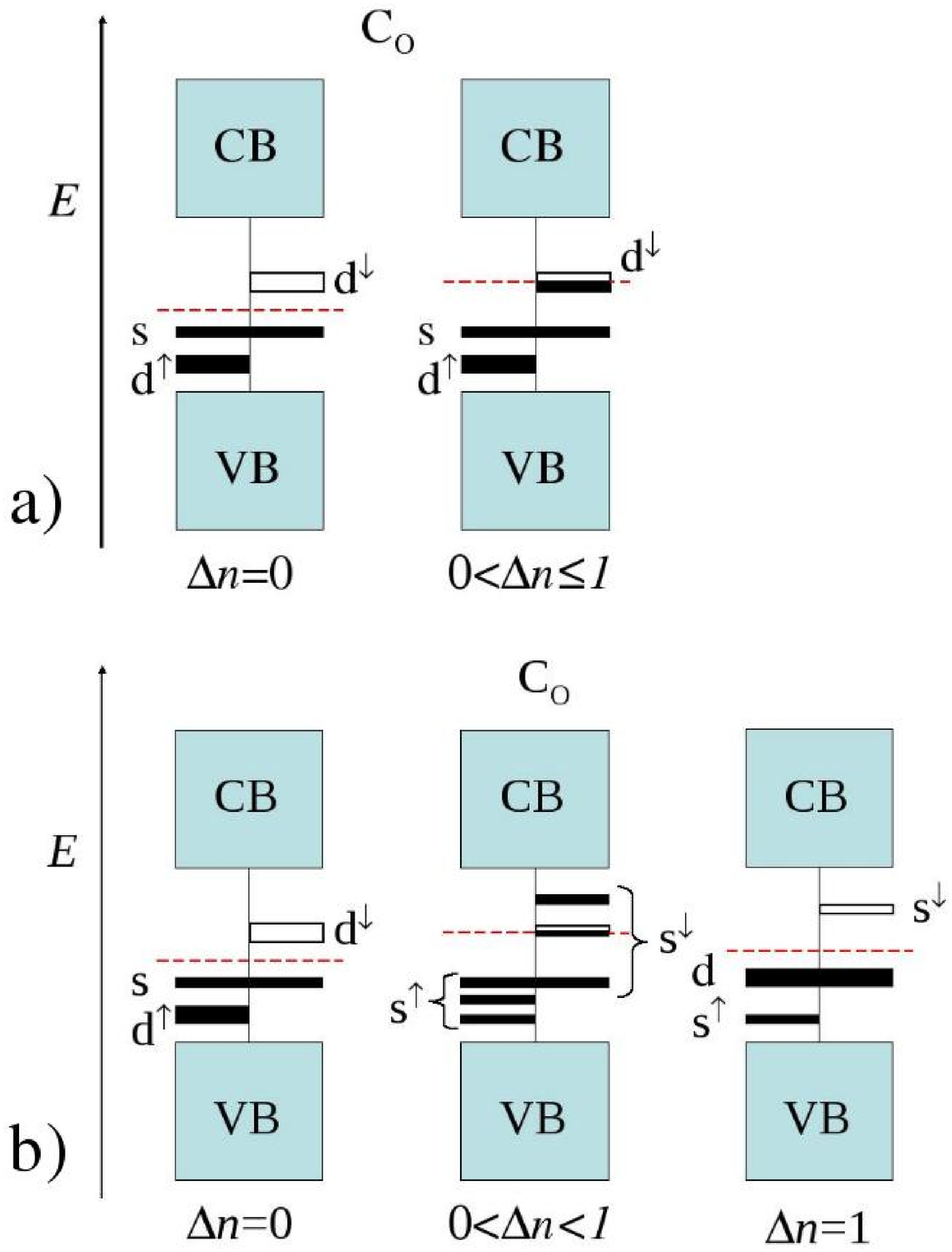}}
\caption{(Color on line) Cartoon representing the level occupation of C$_\mathrm{O}$ upon electron doping. In panel (a) the energy 
levels are those of the local geometry obtained from a relaxation initiated at the C$_\mathrm{O}$ configuration, while in (b)
those obtained from a relaxation initiated at the N$_\mathrm{O}$ configuration. For $\Delta n$=0, in both cases a doubly occupied 
spin-polarized doublet (d$^\uparrow$) is followed by a doubly occupied unpolarized singlet (s) and separated by the Fermi level (red 
dashed line) from its minority spin component (d$^\downarrow$). Upon doping ($\Delta n>0$), in (a) the minority doublet gets progressively 
filled, while in (b) the doublet further splits into two singlets and the additional fractional charge distributes over the lower of
the empty minority singlets. In both cases the magnetic moment goes as $\mu=(2-\Delta n)\mu_\mathrm{B}$.}\label{Fig1}
\end{figure}

The changes in the electronic structure as a function of electron doping are illustrated in the cartoon of Fig.~\ref{Fig1} for the
two obtained geometries. In both cases the additional charge remains localized at the dopant site and the magnetic moment 
varies as $\mu=(2-\Delta n)\mu_\mathrm{B}$. The main difference between the electronic 
structures of the two geometries originates from their different orbital occupation. For the C$_\mathrm{O}$-relaxed structure
the fractional charge progressively occupies the minority empty doublet associated to the four long bonds, spreading 
evenly among them [see Fig.~\ref{Fig1}(a)]. In contrast the $p$-orbitals of the N$_\mathrm{O}$-relaxed structure form 
a set of closely spaced singlets. Hence the additional fractional charge occupies the first of the available empty singlets and 
localized further along the direction of the bond of intermediate length [see Fig.~\ref{Fig1}(b)]. These differences persist
up to $\Delta n$=1 where the C$_\mathrm{O}$-relaxed structure is metallic, while the N$_\mathrm{O}$-relaxed is insulating.

Unfortunately the ASIC functional, although constructs a valuable potential so that atomic relaxation can be carried out, 
does not provide accurate total energies. Therefore we cannot distinguish energetically between the two geometries 
found. We have performed additional ASIC total energy calculations at the un-relaxed C$_\mathrm{O}$ and N$_\mathrm{O}$
geometries, which seems to suggest a crossover between the two at a doping of $\Delta n$=0.5, however also these
calculations are affected by an intrinsic lack of accuracy and therefore the result must be taken with caution.

Given this uncertainty we decided to take a look at the theory for the local vibronic coupling. For manganites the Jahn-Teller distortion 
is driven by the coupling of doubly degenerate $d$-shell $e_g$ states to the only two normal modes having the same symmetry. 
In contrast the $2p$ levels of the N in MgO form a set of degenerate $t_{1u}$  levels, which are coupled with normal modes of symmetry 
$e_g$ and $t_{2g}$. Importantly, in this case the $e_g$ modes tend to stabilize a tetragonal distortion, while the $t_{2g}$ a trigonal 
one resulting in a competition between the two\cite{21}. Furthermore, when anaharmonic corrections are considered, a structure of C$_2$ 
symmetry minimizes the energy. It is notable that the highest occupied level for one hole over the impurity in this geometry is $1/\sqrt{2} 
(p_x \pm p_y)$ and could account for the orbital occupation found for the C$_\mathrm{O}$-relaxed structure at $\Delta n$=1. Unfortunately, 
our relaxations always ends up with a tetragonal distortion and the symmetry around the impurity is never of C$_2$-type. The vibronic theory 
then predicts that the ground state is the one with the hole localized on just one of the three $p$ orbitals. Thus we conclude that simple
symmetry arguments seem to support the relaxation initiated by the N$_\mathrm{O}$ atomic coordinates, i.e. the one converging
to a local geometry having two long, two short and two intermediate bonds. 

After having studied C$_\mathrm{O}$, we briefly take a look at N$_\mathrm{O}$ under doping. This time we find that the unpaired singlet
gradually fills upon electron doping, thus that the distortion gradually reduces to a perfectly cubic symmetry and the moment follows 
$\mu=(1-\Delta n)\mu_\mathrm{B}$. Interestingly we find that the residual hole always localizes over the
longer of the Mg-N bonds and that no magnetic coupling is found for every $\Delta n$, despite the material remains metallic. 

In conclusion, we investigated the effects of electron doping on the magnetic moment formation of N- and C-doped MgO. We find 
that for all doping concentrations the impurity levels are deep in the MgO gap and the magnetic moment is stable against charge 
fluctuations. However, lattice distortion always promotes the localization of the doping hole, thus reducing the chance of long
range magnetic coupling between impurities. 


This work is sponsored by Science Foundation of Ireland. Computational resources have been provided 
by the HEA IITAC project managed by TCHPC and by ICHEC.
{\small{

}}


\begin{thebibliography}{100}
\bibitem{1} T.~Dietl, H.~Ohno, F.~Matsukura, J.~Cibert and D.~Ferrand, Science \textbf{287}, 1019 (2000). 
\bibitem{Behan}A.J.~Behan, A.~Mokhtari, H.J.~Blythe, D.~Score, X.-H.~Xu, J.R.~Neal, A.M.~Fox and G.A.~Gehring,
Phys. Rev. Lett. {\bf 100}, 047206 (2008).
\bibitem{DasZnO}C.D.~Pemmaraju, R.~Hanafin, T.~Archer, H.B.~Braun and S.~Sanvito, Phys. Rev. B {\bf 78}, 054428 (2008).
\bibitem{4}M.~Venkatesan, C.~B.~Fitzgerald and J.~M.~D.~Coey, Nature (London)  {\bf 430}, 630 (2004).
\bibitem{5}S.~D.~Yoon, Y.~Chen, A.~Yang, T.~L.~Goodrich, X.~Zuo, D.~A.~Arena, K.~Ziemer, C.~Vittoria and V.~G.~Harris, J. Phys.: Condens. Matter {\bf 18}, L355 (2006).
\bibitem{6} S.~D.~Yoon, Y.~Chen, A.~Yang, T.~L.~Goodrich, X.~Zuo, K.~Ziemer, C.~Vittoria and V.~G.~Harris, J. Magn. Magn. Mat. {\bf 309}, 171 (2007).\\
S.~D.~Yoon, Y.~Chen, A.~Yang, T.~L.~Goodrich, X.~Zuo, K.~Ziemer, C.~Vittoria and V.~G.~Harris, J. Magn. Magn. Mat. {\bf 320}, 597 (2008).
\bibitem{7} A.K.~Rumaiz, B.~Ali, A.~Ceylan, M.~Boggs. T.~Beebe and S.~I.~Shah, Sol. State Comm. {\bf 144}, 334 (2007)
\bibitem{8} R.~P.~Panguluri, P.~Kharel, C.~Sudakar, R.~Naik, R.~Suryanarayanan, V.~M~Naik, A.~G.~Petukhov, B.~Nadgorny and G.~Lawes, arXiv:0808.1123.  
\bibitem{2}H.~Pan, J.~B.~Yi, L.~Shen, R.~Q.~Wu, J.~H.~Yang, J.~Y.~Lin, Y.~P.~Feng,J.~Ding, L.~H.~Van and J.~H.~Yin, Phys. Rev. Lett. {\bf 99}, 127201 (2007). 
\bibitem{3}S.~Zhou, Q.~Xu, K.~Potzger, G.~Talut, R.~Groetzschel, J.~Fassbender, M.~ Vinnichenko, J.~Grenzer, M.~Helm, H.~Hochmuth, M.~Lorenz, M.~Grundmann and H.~Schmidt, 
Appl. Phys. Lett. {\bf 93}, 232507 (2008).
\bibitem{16} A.~Sundaresan, R.~Bhargavi, N.~Rangarajan, U.~Siddesh and C.~N.~R.~Rao, Phys. Rev. B, {\bf 74}, 161306(R) (2006).
\bibitem{9}Q.~Zhao, P.~Wu, B.~L.~Li, Z.M.~Lu and E.Y.~Jiang, J. Appl. Phys. {\bf 104}, 073911 (2008).
\bibitem{10}K.~Potzger, S.~Zhou, J.~Helm and J.~Fassbender, Appl. Phys. Lett. {\bf 92}, 182504 (2008).
\bibitem{11}C.~D.~Pemmaraju and S.~Sanvito, Phys. Rev. Lett {\bf 94}, 217205 (2005).
\bibitem{13}A.~Droghetti, C.~D.~Pemmaraju and S.~Sanvito, Phys. Rev. B, {\bf 78}, 140404(R) (2008). 
\bibitem{12}G.~Bouzerar and T.~Ziman, Phys. Rev. Lett. {\bf 96}, 207602 (2006).
\bibitem{14}V.~Pardo and W.~E.~Pickett, Phys. Rev. B, {\bf 78}, 134427 (2008).
\bibitem{15}I.~S.~Elfimov, A.~Rusydi, S.~I.~Csiszar, Z.~Hu, H.~H.~ Hsieh, H.~J.~Lin, C.~T.~Chen, R.~Liang and G.~A.~Sawatzky, Phys. Rev. Lett. {\bf 98},137202 (2007).
\bibitem{17}M.~Wang, R.~P.~Campion,A.~W.~Rushforth, K.~W.~Edmounds,C.~T.~Foxon and B.~L.~Gallagher arXiv:0808.1464.
\bibitem{18}E.~Dagotto, T.~Hotta and A.~Moreo, Phys. Rep. {\bf 344}, 1 (2001).
\bibitem{19}J.~M.~Soler, E.~Artacho, J.~D.~Gale, A.~Garc\'{\i}a, J.~Junquera, P.~Ordej\'{o}n and D.~S\'{a}nchez-Portal, J.~Phys.: Condens. Matter {\bf 14}, 2745(2002).
\bibitem{20}C.~D.~Pemmaraju, T.~Archer, D.~Sanchez-Portal and S.~Sanvito, Phys. Rev. B {\bf 75}, 045101 (2007).
\bibitem{21}R.~Englman (1972) {\it The Jahn-Teller Effect In Molecules And Crystals} (John Wiley, London), page 53.
\end{thebibliography}
\end{document}